\documentclass[preprint,nofootinbib,nobibnotes,groupaddress,preprintnumbers,aps,prd]{revtex4}
\usepackage{tipa}
\usepackage{bm,epsfig,mathrsfs,amsmath,amssymb,graphicx}
\usepackage{epsfig,amsmath,graphicx,amssymb}
\usepackage{graphicx}
\usepackage{dcolumn}
\usepackage{bm}
\usepackage[toc,title,titletoc,header]{appendix}

\date{\today}

\begin{document}

\title{Power Fluctuations of An Irreversible Quantum
Otto Engine}
\author{Guangqian Jiao$^{1}$}
\author{Shoubao Zhu$^1$}

\author{Jizhou He$^{1}$}
\author{Yongli Ma$^{2}$}
\author{Jianhui Wang$^{1,2}$}\email{wangjianhui@ncu.edu.cn}

\affiliation{ $^1\,$ Department of Physics, Nanchang University,
Nanchang 330031, China\\ $^2\,$   State Key Laboratory of Surface
Physics and Department of Physics, Fudan University, Shanghai
200433, China}

\begin{abstract}
We derive the general probability distribution function of
stochastic work for quantum Otto engines in which both the isochoric
and driving
 processes are irreversible due to finite time duration.  The time-dependent
power fluctuations, average power, and thermodynamic efficiency  are
explicitly obtained for a complete cycle operating  with an
analytically solvable two-level system. We show that, there is a
trade-off between efficiency (or power) and power fluctuations.

 PACS number(s): 05.70.Ln

\end{abstract}

\maketitle
\date{\today}
\section{Introduction}
The second law of thermodynamics tells us that any heat engines
working between a hot and a cold thermal bath of constant inverse
temperatures $\beta_h$ and $\beta_c$ (with $\beta=1/T$ and
$k_B\equiv1$)  are not able to run more efficiently than a
reversible Carnot cycle with its efficiency
$\eta_C=1-\beta_h/\beta_c$. When  a cyclic heat engine runs at a
Carnot efficiency,  its cycle operation consisting of consecutive
thermodynamic processes  must be reversible   and  thus the power
output becomes null.
 Practically heat engines must proceed in a finite time period  accounting for
 irreversibility
and produce finite power output \cite{1,2,3,4,5,6}. The performance in
finite time was intensively studied  for both quantum\cite{7,8} and
classical\cite{4} engines. For an adequate description of heat
engines, the effects induced by finite-time cycle operation on the
machine performance have to be considered by involving both
heat-transfer and adiabatic processes.

Unlike in   macroscopic systems where both work and heat are
deterministic, the  work \cite{9,10,11,12,13,14,15,16,17} and heat \cite{18,19,20} for
microscale systems are random due to thermal and quantum
fluctuations\cite{8,10,21,22,23,24}. The statistics of either work or
efficiency for heat engines at microscale  were examined
experimentally and theoretically\cite{8,16,22,23,25,26,27,28,29,30,31,32,33,34}, but under
the assumption that either adiabatic\cite{8,23} or isothermal
strokes\cite{22} are reversible.  To our knowledge, the power
fluctuations have not been examined so far for an irreversible
quantum heat engine where the isothermal and adiabatic strokes are
finite-time and thus both of them are away from  reversible limit.

In this paper,  we  derive the general expression for the
distribution function of quantum work and heat  along irreversible
quantum Otto engines composed of  two finite-time isochoric and two
driving strokes\cite{8,22,23,34,35,36,37,38}. This distribution function
allows us to obtain the quantum work statistics explicitly depending
on the time evolution dynamics of the  two isochoric and two driving
processes. We then analytically examine the power statistics of an
Otto cycle working with an exactly, solvable two-level system. The
effects of irreversibility induced by finite-time duration of either
driving or isochoric strokes on machine performance and power
statistics are discussed. We finally demonstrate that
irreversibility results in deterioration of efficiency and power but
increasing the power fluctuations, thereby indicating that the price
for improving machine efficiency is increasing power fluctuations.

\section{Model}

\begin{figure}[tb]
\centering
\includegraphics[width=2.8in]{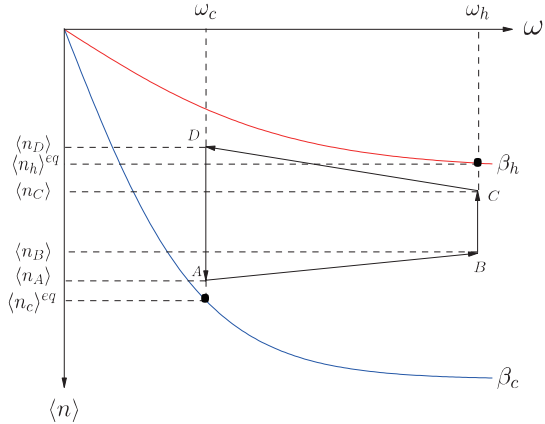}
\caption{Schematic diagram of a two-level quantum Otto   cycle operating
between a hot and a cold thermal bath of inverse constant inverse
temperatures ($\beta_h$ and $\beta_c$) in $(\omega,\langle
n\rangle)$ plane. The cycle consists of the two driving strokes
(connecting states $A$ and $B$, and $C$ and $D$), where the system
isolated from the two thermal baths  evolves unitary transformation
and the respective time durations are $\tau_{ch}$ and $\tau_{hc}$,
and two isochoric strokes (connecting states $B$ and $C$, and $D$
and $A$), where the system is kept in thermal contact with the hot
and the cold reservoir, respectively, and these respective time
durations are $\tau_h$ and $\tau_c$. The average populations at the
four instants $A, B, C, D$ can be expressed as $\langle
n_A\rangle=\langle n(0)\rangle, \langle n_B\rangle=\langle
n(\tau_{ch})\rangle, \langle n_C\rangle= \langle
n(\tau_{ch}+\tau_h)\rangle, \langle n_D\rangle = \langle
n(\tau_{cyc}-\tau_c)\rangle$. The mean population $\langle
n_A\rangle (\langle n_C\rangle)$ at the end of the cold (hot)
isochoric stroke would approach its asymptotic value $\langle
n_c\rangle^{eq}$ ($\langle n_h\rangle^{eq}$), when and only when the
thermalization is completed after infinite long duration of the
system-bath interaction interval $\tau_{c}$ ($\tau_h$).}
\label{engine}
\end{figure}

 The model of a quantum Otto engine is sketched in Fig. \ref{engine}. This
machine consists of two isochoric branches, one with a hot and
another with a cold thermal bath where the system frequency is kept
constant, and two driving strokes, where the system (which is
isolated from the two heat reservoirs and driven by the external
field) undergoes unitary transformation. The four branches can be
described as follows.

During the adiabatic compression $A\rightarrow B$, the system is
isolated from the two thermal baths but driven by the external
field, and it undergoes a unitary expansion from time $t=0$ to
$t=\tau_{ch}$. Initially, the system of inverse temperature
$\beta_A$ is assumed to be (local) thermal equilibrium \cite{2,3},
which means that the thermal occupation probability at instant $A$
takes the canonical form
\begin{equation}
 p_{n}^{0}=\frac{e^{-\beta_{A}E^{c}_{n}}}{Z_{A}} \label{pn0},
 \end{equation}
with the partition function  $Z_{A}=\sum_n e^{-\beta_A E_n^c }$.
   The transition probability from eigenstate $|n\rangle$ to
$|m\rangle$ is given by
\begin{equation}
 p_{n\rightarrow m}^{\tau_{ch}}=|\langle
 n|U_{\mathrm{com}}|m\rangle|^2, \label{pnm}
\end{equation}
where $U_{\mathrm{com}}$ denotes  the unitary time evolution
operator along the compression.  In the special case when the system
Hamiltonian evolves slowly enough by satisfying quantum adiabatic
condition \cite{38}, the system remains in the same state and the
transition probability becomes $p_{n\rightarrow
m}^{\tau_{ch}}=\delta_{nm}$, with the Kronecker delta function
$\delta$.

 The next step is   the hot isochore
$B\rightarrow C$, where the system with constant frequency
$\omega=\omega_h$ is in contact with
 the hot thermal bath in  a time duration $\tau_h$. The probability
density of the stochastic heat absorbed from the hot bath  can be
determined by the conditional probability to obtain
\begin{equation}
p(q_h)=\sum_{k,l}\delta[q_h-(E_{l}^{h}-E_{k}^{h})]p_{k\rightarrow
l}^{\tau_h}.    \label{pqpk}
\end{equation}
After this system-bath interaction interval, the system is assumed
to be at the local thermal equilibrium state of inverse temperature
$\beta_C(\ge\beta_h)$, which means that $p_{k\rightarrow
l}^{\tau_h}={e^{-\beta_{C}E^{h}_{l}}}/{Z_{C}}$ with partition
function $Z_{C}=\sum_l {e^{-\beta_{C}E^{h}_{l}}}$.

During the expansion $C\rightarrow D$, the system is isolated from
 these two thermal baths in a time period $\tau_{hc}$ while its
frequency changes back to $\omega_c$  from $\omega_h$. Like in the
driving compression, there is a transition probability from initial
state $i$ to final one $j$, which reads
\begin{equation}
p_{i\rightarrow j}^{\tau_{hc}}=|\langle
i|U_{\mathrm{exp}}|j\rangle|^2 \label{pij},
\end{equation}
where  $U_{\mathrm{exp}}$ is the time evolution operator
 along the expansion. At the initial instant $C$
 the system approaches the thermal equilibrium and the
 occupation probabilities take the form $p_i^{\tau_{ch}+\tau_h}=\delta_{il}p_{k\rightarrow
 l}^{\tau_h}$, where $p_{k\rightarrow
 l}^{\tau_h}$ was defined in Eq. (\ref{pqpk}).

On the fourth branch $D\rightarrow A$,  the system with constant
frequency $\omega=\omega_c$  is coupled to a cold reservoir of
inverse temperature $\beta_c$ in a time period $\tau_{c}$. After a
cycle with the total period
$\tau_{cyc}=\tau_{ch}+\tau_h+\tau_{hc}+\tau_c$ the system returns to
its initial state $A$, we
  can easily determine the  quantum heat $q_c$
 similar to that of the quantum heat $q_h$. As emphasized, at instant $A$ the
 system tends to be (local) thermal equilibrium after system-bath interaction interval,
 which is consistent with the assumption of canonical form Eq. (\ref{pn0}).

In view of the fact that  the work per cycle is produced only in the
two driving strokes $A\rightarrow B$ and $C\rightarrow D$, the
quantum work output can be obtained by determining the total work
produced along these two microscopic trajectories to arrive at
\begin{equation}
\mathrm{w}[n(0)\rightarrow m(\tau_{ch});
i(\tau_{ch}+\tau_h)\rightarrow
j(\tau_{cyc}-\tau_c)]=(E_i^h-E_j^c)-(E_m^h-E_n^c), \label{stwork}
\end{equation}
where  $E_{n}^{c}$ and $E_{m}^{h}$  ($E_i^h $ and $ E_j^c$) denote
the respective energy eigenvalues at the initial and final instants
of the compression (expansion). Here $n(t)$ are quantum numbers
indicating the states occupied by the system at time $t$.

The states $|n(0)\rangle$ and $| m(\tau_{ch})\rangle$ can be assumed
to be independent  of $i(\tau_{ch}+\tau_h)$ and
$j(\tau_{cyc}-\tau_{c})$,  since the system would relax to thermal
equilibrium in an isochoric process if its time duration is long
enough. This leads to the main result of this paper, namely, the
following expression for the probability density of the work
$\mathrm{w}$:
\begin{equation}
p(\mathrm{w})=\sum_{m,n,i,j}p_{n\rightarrow m}^{\tau_{ch}}
p_n^0p_{i\rightarrow
j}^{\tau_{hc}}p_i^{\tau_{ch}+\tau_h}\delta\{\mathrm{w}-\mathrm{w}[n(0)\rightarrow
m(\tau_{ch});  i(\tau_{ch}+\tau_h)\rightarrow
j(\tau_{cyc}-\tau_c)]\}, \label{pw0}
\end{equation}
where $p_{n\rightarrow m}^{\tau_{ch}}$ and $p_{i\rightarrow
j}^{\tau_{hc}}$ were defined in Eq. (\ref{pnm}) and Eq. (\ref{pij}), respectively. The
probability distribution function Eq. (\ref{pw0}) for the irreversible Otto
engine allows us to determine all moments of quantum work: $\langle
\mathrm{w}^k\rangle=\int \mathrm{w}^k p(\mathrm{w}) d \mathrm{w}$
($k=1,2,\dots).$ This result reduces to that  previously obtained in
a quasistatic \cite{27} or an endoreversible \cite{8} Otto engine
when $\xi\rightarrow 0$ or $\tau_{c,h}\rightarrow\infty$. We present
it here in a broader context by arguing that irreversibility is
unavoidable in a finite-time cyclic engine where both the driving and
system-bath interaction steps  are away from quasistatic limit. For
an endoreversible model where the two  driving strokes  are
isentropic, we recover the work fluctuations \cite{8} by setting
$p_{n\rightarrow m}^{\tau_{ch}}=\delta_{nm}$ and $p_{i\rightarrow
j}^{\tau_{hc}}=\delta_{ij}$,
$p(\mathrm{w})=\sum_{n,i}p_n^0p_i^{\tau_{ch}+\tau_h}\delta\{\mathrm{w}-\mathrm{w}[n(0);
i(\tau_{ch}+\tau_h)]\}$.
\section{Quantum Otto engine working with a two-level system}
We now consider a quantum Otto engine that works with a two-level
system of the eigenenergies $E_+=\hbar\omega/2$ and
$E_-=-\hbar\omega/2$. As  the occupation probabilities at these two
eigenstates $p_+=e^{-\beta\hbar\omega_c/2}/Z_A$ and
$p_-=e^{\beta\hbar\omega_c/2}/Z_A$, where the partition function $
Z_A=e^{-\beta_A\hbar\omega_c/2}+e^{\beta_A\hbar\omega_c/2}=2\cosh\left(\frac{\beta_A\hbar\omega_c}2\right),
$  the mean population at instant $A$ (with time $t=0$) can be
determined  by using $\langle n\rangle=\sum_n n p_n$ to arrive at
\begin{equation}
\langle
n(0)\rangle=-\frac{1}2\tanh\left(\frac{\beta_A\hbar\omega_c}2\right).
 \label{n0n2}
\end{equation}

 The mean
population at final instant $B$ can then be determined according to
$\langle n(\tau_{ch})\rangle=\sum_{n,m} n p_{m\rightarrow
n}^{\tau_{ch}}p_m^0$, which together with Eqs. (\ref{pn0}) and (\ref{pnm}), leads  to
\begin{equation}
\langle n(\tau_{ch})\rangle=\left(1-2\xi\right)\langle n(0)\rangle,
\label{nbna}
\end{equation}
where $\xi=|\langle \pm|U_{\mathrm{com}}|\mp\rangle|^2$.  Using
$\langle n(\tau_{cyc}-\tau_{c})\rangle=\sum_{n,m} n p_{m\rightarrow
n}^{\tau_{hc}}p_m^{\tau_{ch}+\tau_{h}}$, for the unitary expansion
$C\rightarrow D$ there is the relation:
\begin{equation}
\langle n(\tau_{cyc}-\tau_{c})\rangle=(1-2\xi)\langle
n(\tau_{ch}+\tau_h)\rangle, \label{ndnc}
\end{equation}
where we have assumed $\xi=|\langle
\pm|U_{\mathrm{exp}}|\mp\rangle|^2=|\langle
\pm|U_{\mathrm{com}}|\mp\rangle|^2$. $\xi$ is called the
adiabaticity parameter representing the probability of transition
between state $|+\rangle$ and $|-\rangle$ during the compression or
expansion, and the probability of no state transition
 is accordingly $|\langle
\pm|U_{\mathrm{exp}}|\pm\rangle|^2=|\langle
\pm|U_{\mathrm{com}}|\pm\rangle|^2=1-\xi$. Since the population
$\langle n\rangle$ at any instant along the cycle is negative,
 $\xi$ must be situated between $0\le\xi<1/2$. The
adiabaticity parameter $\xi$ depends on the speed at which the
driving process is performed\cite{7,22,36}, and it is thus a
monotonic decreasing function of time duration $\tau_{h,c}$. Unlike
in an adiabatic phase where the time scale of state change must be
much larger than the dynamical one and the probability of state
change is vanishing ($\xi=0$), during the nonadiabatic
  driving  stroke the rapid change in frequency $\omega$
 leads to inner friction and  results in possible state transitions
($\xi>0$)\cite{22,36,39,40,41,42}. Such  nonadiabatic internal
dissipation accounts for irreversible entropy production and an
increase in mean population $\langle n\rangle$ (see Fig. \ref{engine}).

We are interested in the finite-time operation of the Otto engine in
which the isochoric processes are far away from quasistatic limit
and complete thermalization can thus not be achieved for the system.
Using the master equation of stochastic thermodynamics 
, one can find that the mean populations
$\langle n(0)\rangle$ and $\langle n(\tau_{ch}+\tau_h)\rangle$  can
be expressed in terms of the corresponding asymptotic equilibrium
values
 $\langle n_c\rangle^{eq}$ and $\langle n_h\rangle^{eq}$ (see Appendix \ref{A}),
\begin{equation}
\langle n(0)\rangle=\langle n_c\rangle^{eq}+[\langle
n(\tau_{cyc})\rangle-\langle n_c\rangle^{eq}]e^{-\gamma_c\tau_c}, \label{ncc}
\end{equation}
\begin{equation}
\langle n(\tau_{ch}+\tau_{h})\rangle=\langle n_h\rangle^{eq}+[\langle
n(\tau_{ch})\rangle-\langle n_h\rangle^{eq}]e^{-\gamma_h\tau_h}. \label{nah}
\end{equation}
Using Eqs. (\ref{nbna}),(\ref{ndnc}),(\ref{ncc}) and (\ref{nah}), $\langle n(0)\rangle$ and $\langle n(\tau_{ch}+\tau_{h})\rangle$ can be rewritten as
\begin{equation}
\langle n(0)\rangle=\langle n_c \rangle^{eq}+\Delta_c, \langle
n(\tau_{ch}+\tau_h)\rangle=\langle n_h \rangle^{eq}+\Delta_h
 \end{equation}
where
\begin{equation}
\Delta_h=\frac{\left(2\xi-1\right)\left[\left(2\xi-1\right)\langle
n_h\rangle^{eq}+\langle n_c\rangle^{eq}\right]-x_c\left[\langle
n_h\rangle^{eq}+\left(2\xi-1\right)\langle n_c\rangle^{eq}\right]}{x_hx_c-\left(2\xi-1\right)^2},
\end{equation}
\begin{equation}
\Delta_c=\frac{\left(2\xi-1\right)\left[\left(2\xi-1\right)\langle
n_c\rangle^{eq}+\langle n_h\rangle^{eq}\right]-x_h\left[\langle
n_c\rangle^{eq}+\left(2\xi-1\right)\langle n_h\rangle^{eq}\right]}
{x_hx_c-\left(2\xi-1\right)^2}.
\end{equation}
Here  $x_h\equiv e^{\gamma_h\tau_h}$ and $x_c\equiv
e^{\gamma_c\tau_c}$ denote the effective time durations along the
hot and cold isochoric branches, respectively. Here $\langle
n_c\rangle^{eq}$ and $\langle n_h\rangle^{eq}$  are  achieved in
the reversible, quasistatic limit when  $x_h, x_c \rightarrow\infty$
leads to $\Delta_{h,c}\rightarrow0$, whether $\xi$ is zero or not.
However, $\Delta_c$ and $\Delta_h$ are still positive for finite
values of $x_c$ and $x_h$ even for two reversible driving processes
 with $\xi\rightarrow 0$. These corrections $\Delta_c$ and $\Delta_h$
indicate that how far the heat-transfer processes deviate from the
reversible limit, and imply that irreversibility  is exclusively
caused by heat transferred between the system and the thermal bath.
 Such a deviation  is quite natural in both  quantum heat engines
and classical context when the heat-transfer processes are
irreversible due to finite-time duration.

For this two-level engine, the probability density of quantum work
Eq. (\ref{pw0}) can be analytically obtained as,
\begin{equation}
\begin{aligned}
p(\mathrm{w})=&\left[\frac{1}{2}+2\langle n(0)\rangle \langle
n(\tau_{ch}+\tau_h)\left(1-2\xi\right)-\left(1-\xi\right)\xi\right]\delta(\mathrm{w})\\
&+\frac{1}{2}\left[\left(1-2\langle
n(0)\rangle\right)\left(1-\xi\right)\xi\right]\delta(\mathrm{w}+\hbar\omega_c)\\
&+\frac{1}{2}\left[\left(2\langle
n(\tau_{ch}+\tau_h)\rangle+1\right)\left(1-\xi\right)\xi\right]\delta(\mathrm{w}-\hbar\omega_h)\\
&+\frac{1}{4}\left[\left(2\langle
n(\tau_{ch}+\tau_h)\rangle-1\right)\left(2\langle
n(0)\rangle-1\right)\xi^2\right]\delta(\mathrm{w}+\hbar\omega_h+\hbar\omega_c)\\
&+\frac{1}{4}\left[1-\left(2\langle
n(\tau_{ch}+\tau_h)\rangle\right)\left(2\langle
n(0)\rangle+1\right)\left(1-\xi\right)^2\right]\delta(\mathrm{w}+\hbar\omega_h-\hbar\omega_c)\\
&+\frac{1}{4}\left[\left(2\langle
n(\tau_{ch}+\tau_h)\rangle+1\right)\left(1-2\langle
n(0)\rangle\right)\left(1-\xi\right)^2\right]\delta(\mathrm{w}+\hbar\omega_c-\hbar\omega_h)\\
&+\frac{1}{2}\left[\left(2\langle
n(0)\rangle+1\right)\left(1-\xi\right)\xi\right]\delta(\mathrm{w}-\hbar\omega_c)\\
&+\frac{1}{4}\left[\left(2\langle n(0)\rangle+1\right)\left(2\langle
n(\tau_{ch}+\tau_h)\rangle+1\right)\xi^2\right]\delta(\mathrm{w}-\hbar\omega_c-\hbar\omega_h)\\
&+\frac{1}{2}\left[\left(1-2\langle
n(\tau_{ch}+\tau_h)\rangle\right)\left(1-\xi\right)\xi\right]\delta(\mathrm{w}+\hbar\omega_h).\label{pw1}
\end{aligned}
\end{equation}

\begin{figure}[tb]
\centering
\includegraphics[width=2.8in]{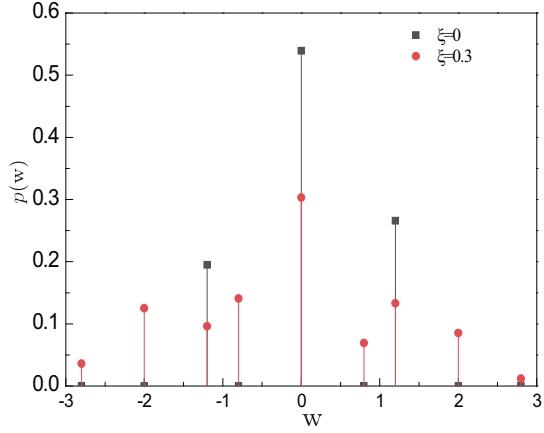}
\caption{The probability distribution $p(\mathrm{w})$ of the quantum work for
the adiabatic (with $\xi=0,$ black squares) and nonadiabatic (with
$\xi=0.3,$ red dots) steps. The parameters are   $x_{c}=x_{h}=8$,
$\omega_c=0.4\omega_h=0.4$, $\beta_h=0.2\beta_c=0.2$, and $\hbar=2$.}
\label{pw}
\end{figure}

The stochastic work
  can take nine different discrete values as shown Fig. \ref{pw}.  The following should be noted in respect of the stochastic work
per cycle: (i) $\mathrm{w}=0$ indicates that the stochastic work by
the system along the expansion is fully counterbalanced by that
during the compression.  (ii)
$\mathrm{w}=\hbar\omega_c-\hbar\omega_h$
($\mathrm{w}=\hbar\omega_h-\hbar\omega_c$)  corresponds to the case
when the system jumps down from  high-energy (low-energy) state to
low-energy (high-energy) one along the hot isochore, namely,
$n=m=1/2$ but $i=j=-1/2$ ( $n=m=-1/2$ and $i=j=1/2$) in Eq. (\ref{pw0}).
These values exist in the adiabatic or nonadiabatic driving. (iii)
There are more values of stochastic work in the nonadiabatic driving
than in the adiabatic case due to quantum fluctuations induced by
nonadiabatic transitions. (iv) Finally the distribution
$p(\mathrm{w})$ is expected to be normalized to one either for
adiabatic or nonadiabatic driving.

Using Eq. (\ref{pqpk}), the average heat injection,
$\langle q_h\rangle=\int q_h p(q_h)dq_h$, can be obtained as
\begin{equation}
\langle q_h\rangle=\hbar\omega_h[\langle n(\tau_{ch}+\tau_h)\rangle
-(1-2\xi)\langle n(0)\rangle ]. \label{qh0}
\end{equation}
By using simple algebra (see Appendix \ref{B} for details),
the average work ($\langle \mathrm{w}\rangle$) and the work
fluctuations ($\delta \mathrm{w}^2=\langle
\mathrm{w}^2\rangle-\langle \mathrm{w}\rangle^2$) can be obtained as
\begin{equation}
\langle \mathrm{w}\rangle=\hbar[\omega_c-(1-2\xi)\omega_h]\langle
n(0)\rangle+\hbar[\omega_h-(1-2\xi)\omega_c]\langle
 n(\tau_{ch}+\tau_h)\rangle,
\label{work}
\end{equation}
and
\begin{equation}
\begin{aligned}
\delta \mathrm{w}^2=&\hbar^2 \omega_h^2 [\frac{1}2 -\langle
n(0)\rangle^2(1-2\xi)^2 - \langle n(\tau_{ch}+\tau_h)\rangle^2]\\
&+\hbar^2 \omega_c^2 [\frac{1}2 -\langle n(0)\rangle^2 -(1-2\xi)^2
\langle n(\tau_{ch}+\tau_h)\rangle^2 ]\\
&-\hbar^2\omega_c\omega_h(1-2\xi) [1-2\langle n(0)\rangle^2  -
2\langle n(\tau_{ch}+\tau_h)\rangle^2].\label{deltaw2}
\end{aligned}
\end{equation}

From Eqs. (\ref{qh0}) and (\ref{work}), the machine efficiency defined by
$\eta=\langle \mathrm{w}\rangle/\langle q_h\rangle$, can be
expressed as
\begin{equation}
 \eta=1-\frac{\omega_c}{\omega_h}\frac{\langle
n(0)\rangle-(1-2\xi)\langle
 n(\tau_{ch}+\tau_h)\rangle}{(1-2\xi)\langle
n(0)\rangle-\langle
 n(\tau_{ch}+\tau_h)\rangle}.\label{eta}
\end{equation}

In case both isochoric and driving processes proceed in finite time,
internal dissipation and uncomplete thermalization occur in the
system resulting in the thermodynamic efficiency Eq. (\ref{eta}) that depends
on the time evolution along each cycle except if these four
processes may be done infinitely long, making the efficiency reduce
to the one for cycles with complete thermalization, $
 \eta=1-\frac{\omega_c}{\omega_h}\frac{(1-2\xi)\langle
 n_h\rangle^{eq}-\langle
n_c\rangle^{eq}}{\langle
 n_h\rangle^{eq}-(1-2\xi)\langle n_c\rangle^{eq}}
$\cite{7,34,36}, or the one for models without internal
dissipation \cite{2,3,35}, $\eta=1-\frac{\omega_c}{\omega_h}$.
\begin{figure}[tb]
\includegraphics[width=2.8in]{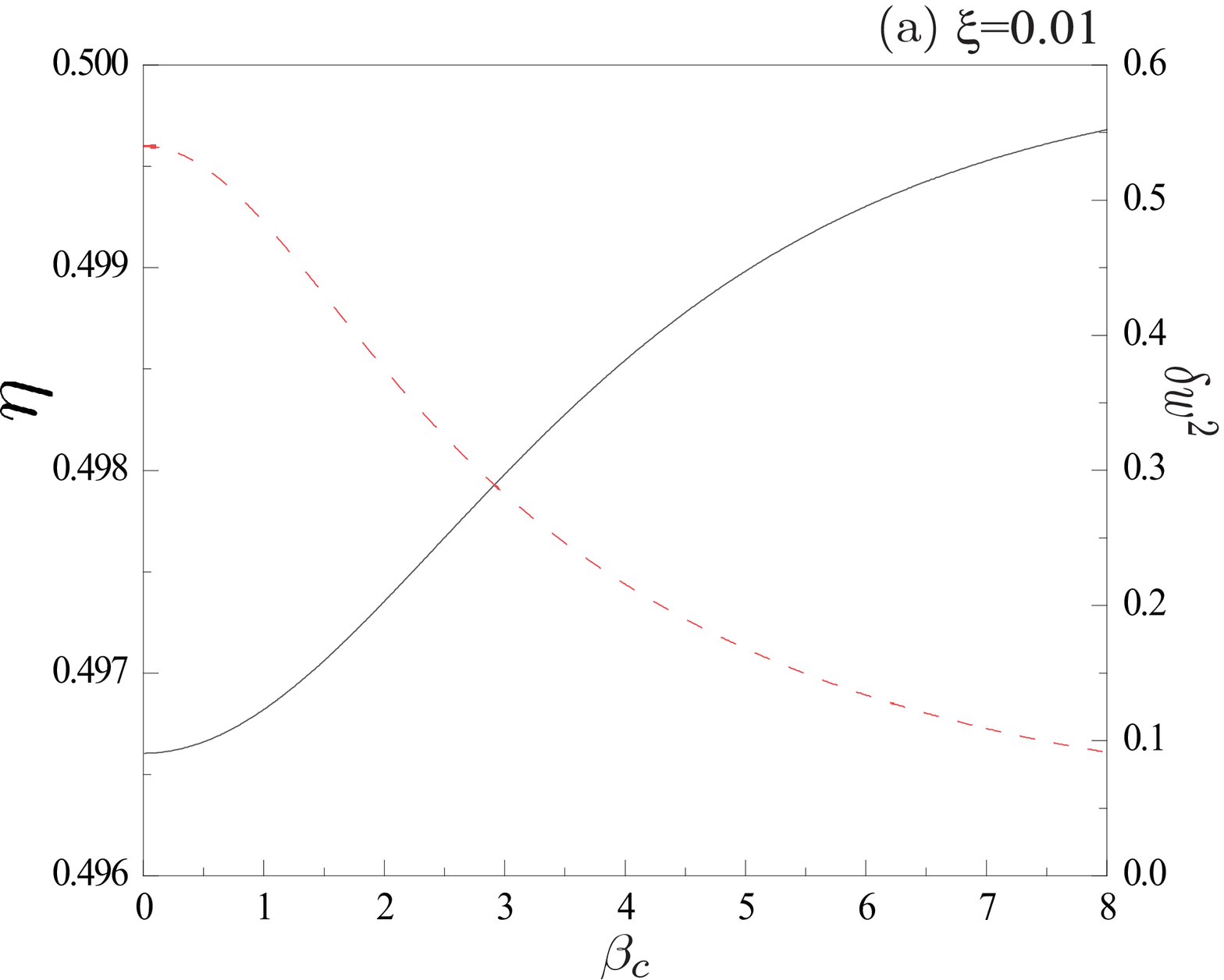}\label{fweta01}
\includegraphics[width=2.8in]{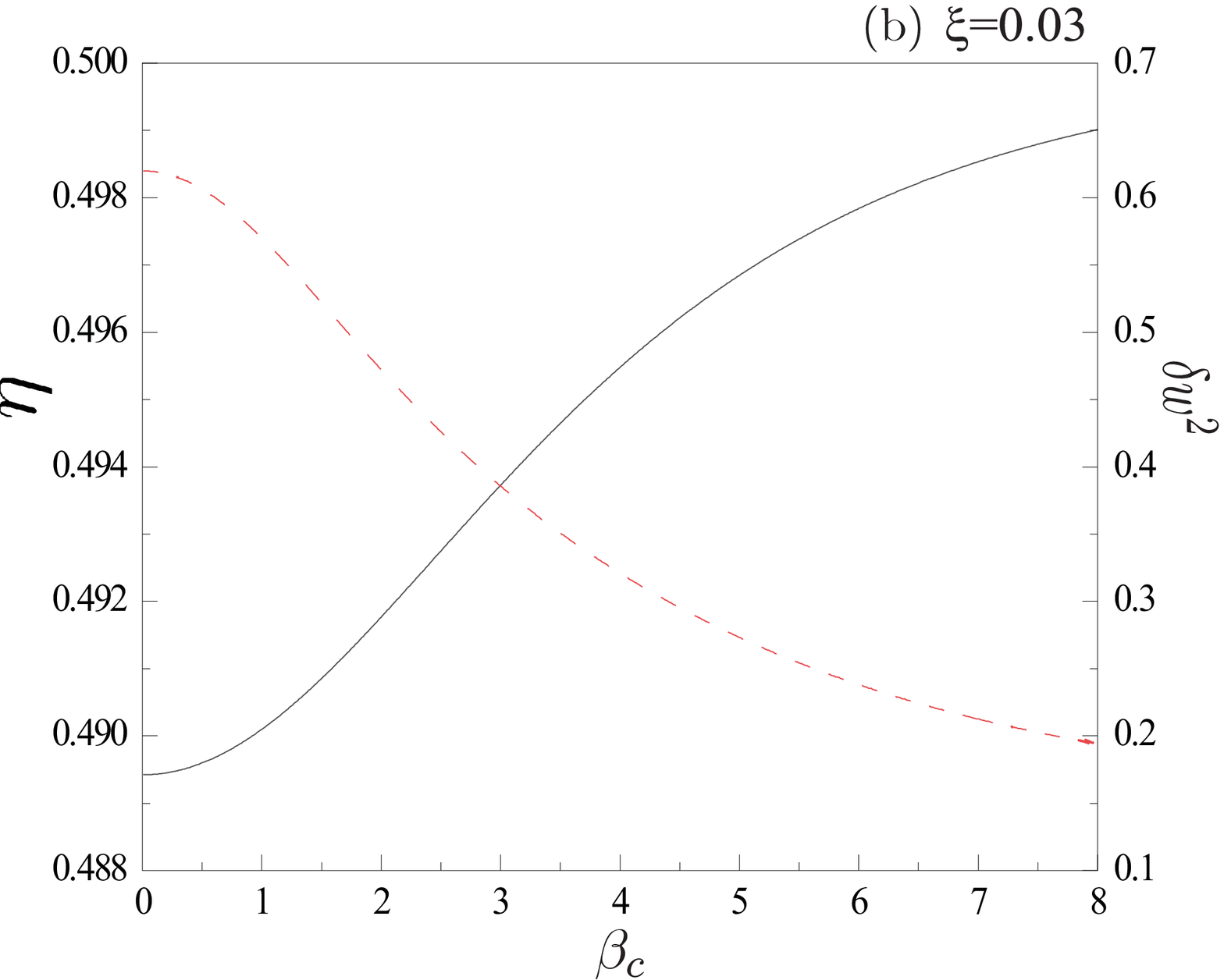}\label{fweta03}
\caption{(Color online) The work fluctuations $\delta \mathrm{w}^2$ (black
solid line ) and the efficiency
 $\eta$ (red dashed line) as a function of the inverse  temperature of cold
 bath $\beta_c=5\beta_h$ for  $\xi=0.01$ (a) and $\xi=0.03$ (b).
The parameters are $x_{c}=x_{h}=8$, $\omega_h=2\omega_c=1$, and
$\hbar=2$.}
\label{fweta}
\end{figure}
The efficiency and work fluctuations as a
function of the inverse temperature $\beta_c$ of cold bath is shown
in Fig. \ref{fweta}(a) and Fig. \ref{fweta}(b). When increasing  temperature, the
efficiency $\eta$ increases but the work fluctuations $\delta
\mathrm{w}^2$ decrease. The efficiency at high temperatures is
larger than that at low temperatures,  showing that the quantum
effects which are of significance in low temperature regime can
improve the machine efficiency. Because the quantum fluctuations
characterizing the low temperature domain are smaller than the
thermal fluctuations dominating   the high temperature region, the
work fluctuations $\delta \mathrm{w}^2$ get increased while the
temperature is increased and \emph{vice versa}. The increase in the
adiabacity parameter $\xi$ yields in a decrease (an increase) in the
machine efficiency (work fluctuations) as it should, thereby showing
that there is a trade-off between efficiency and work fluctuations.

\begin{figure}[tb]
\includegraphics[width=2.8in]{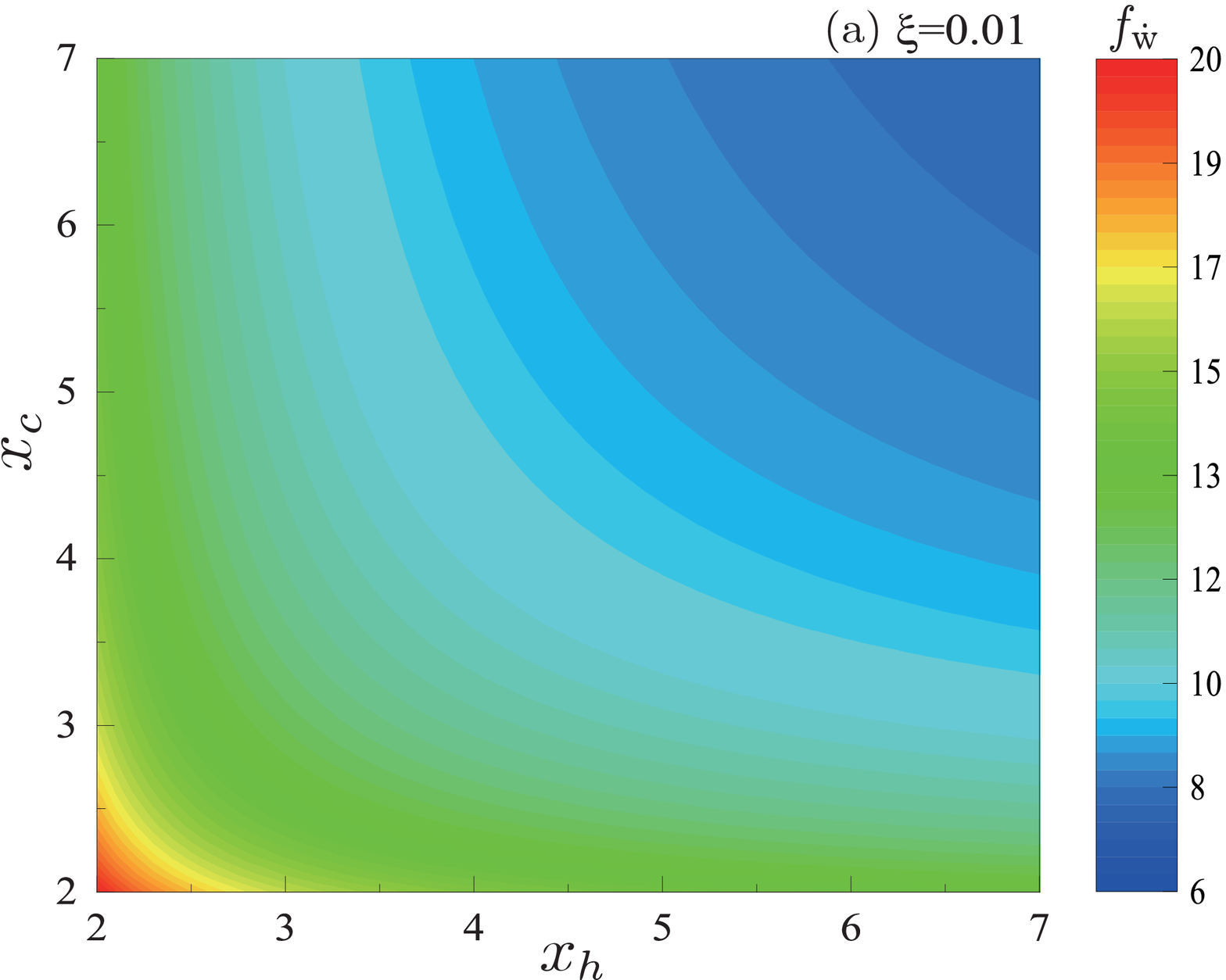}
\includegraphics[width=2.8in]{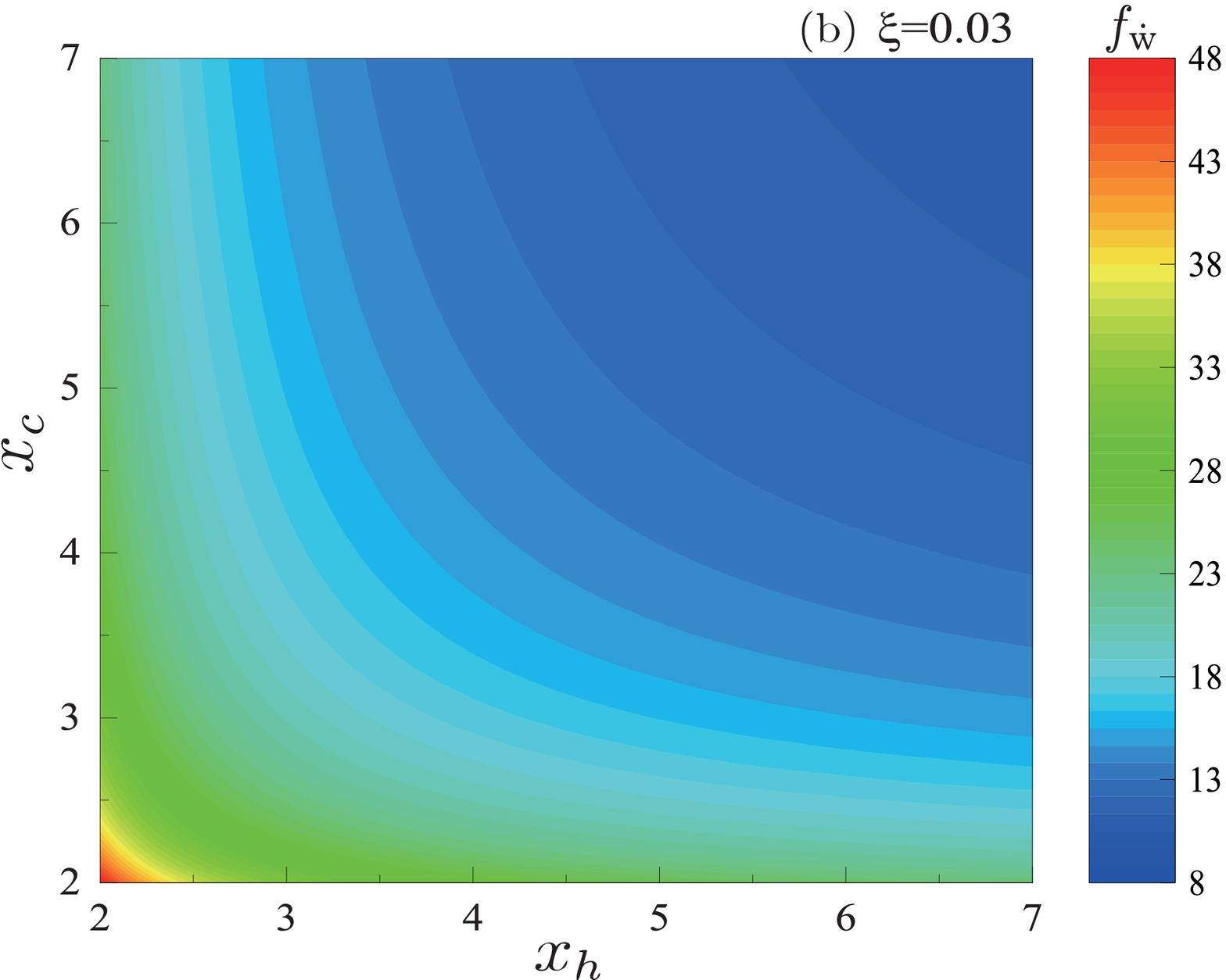}
 \caption{Contour plot of the relative fluctuations of the power
$f_{\dot{\mathrm{w}}}$ in the effective time duration
$(x_{h},x_{c})$ plane for a nonadiabatic  driving, with $\xi=0.01$
($a$) and $\xi=0.03$ ($b$).
  The values of the parameters are $\beta_h=0.2\beta_c=0.2$,  $\omega_h=2\omega_c=1$, and
$\hbar=2$.  }\label{fpx}
\end{figure}

Since the stochastic power  is the work divided by the cycle
 period $\tau_{cyc}$, namely,
$\dot{\mathrm{w}}[|n(0)\rangle; | n(\tau_{ch}+\tau_{h})\rangle] =
\mathrm{w}[|n(0)\rangle; | n(\tau_{ch}+\tau_{h})\rangle]/
\tau_{cyc}$,
 the relative fluctuations of the power $f_{\dot{\mathrm{w}}}$  are equivalent to
corresponding those of work $f_\mathrm{w}$, leading to
$f_{\dot{\mathrm{w}}}=f_\mathrm{w}=\sqrt{\delta
\mathrm{w}^2}/\langle \mathrm{w}\rangle$.  For nonadiabatic driving branches, the relative power fluctuations are
increasing with decreasing effective time durations $x_c$ and $x_h$,
see Fig. \ref{fpx}(a) and Fig. \ref{fpx}(b). Comparison between these two
figures shows that irreversible nonadiabatic friction related to
$\xi$, which is a monotonic decreasing function of $\tau_{h}$ as
well as $\tau_{c}$, leads to an increase in relative power
fluctuations.  This can be understood by the fact that
irreversibility either induced by finite heat flux between the
system and the bath or caused by internal irreversible dissipation
yields larger fluctuations than those in the reversible cycle
operation. In view of the fact that the power is a monotonic
decreasing function of time durations $\tau_h$ and $\tau_c$, we see
that the price one should pay for increasing power is increasing
 power fluctuations.

\section{Conclusions}
In summary, we derived  the probability distribution of stochastic
work of quantum Otto engines working within a cycle period of finite
time that leads to irreversibility in both the two isochoric and two
driving processes. Employing a two-level system as the working
substance of these engines, we find  that, although the average work
is positive, the quantum work may be negative due to quantum
indeterminacy. Afterwards, we derived the analytical expressions for
the power output, efficiency, and work fluctuations, all of which
are dependent on the time allocations the four thermodynamic
processes. We finally determined statistics of work as well as power
at any finite temperatures, and showed that there is a trade-off
between efficiency (or power) and power fluctuations.

\begin{appendices}
\numberwithin{equation}{section}
\section{Time evolution of population along an isochoric process}\label{A}

The dynamics of the system with energy quantization along the
system-bath interaction interval can be  described  by changes in
the occupation probabilities $p_n$ at states $n=0,1,2,\cdots$. This
reduced description in which the dynamical response of the heat
reservoir is cast in kinetic terms.  The master equation is given by
\cite{21,43}
\begin{equation}
\dot{p}_n=\sum_{n'}W_{n,n'}p_{n'}, \label{1}
\end{equation}
where the transition rate matrix $W_{n,n'}$ must satisfy $\sum_n
W_{n,n'}=0$. For the system in contact with a heat bath of constant
temperature $\beta$,  we assume that the transition rates from state
$n'$ to $n$, $W_{nn'}$,  fulfill the detailed balance
$W_{nn'}e^{-\beta E_{n'}}=W_{n'n}e^{-\beta E_n}$, ensuring that the
system can achieve asymptotically the thermal state after an
infinitely long system-bath interaction duration.  At thermal state,
the occupation probabilities $p_n$  achieve their asymptotic
stationary values $\pi_n$. These  $\pi_n$  can be  determined from
the steady-state solution of Eq. (\ref{1}) and given by the Boltzmann
distribution: $ \pi_n(\beta)={e^{-\beta E_n}}/{Z}, $ where $Z=\sum_n
e^{-\beta E_n}$  is the canonical partition function.

For the two-level system where the energy spectrum reads
$E_+=\frac{1}2\hbar\omega$ and $E_-=-\frac{1}2\hbar\omega$, the
master equation Eq. (\ref{1}) becomes
\begin{equation}
  \left(
      \begin{array}{c}
        \dot{p}_+  \\
       \dot{p}_-
      \end{array}
    \right)=\left(
      \begin{array}{cc}
        -W_{-+} & W_{+-}  \\
       W_{-+} & -W_{+-}
      \end{array}  \right)\left(
      \begin{array}{c}
        {p}_+  \\
       {p}_-
      \end{array}
    \right), \label{r2ow}
\end{equation}
where these two transition rates $W_{-+}$ and $W_{+-}$ obey the
detailed balance, $W_{-+}/W_{+-}=e^{-\beta\hbar\omega}$. From Eq. (\ref{r2ow}),  the
 motion for the average population can be obtained as
 \begin{equation}
 \langle \dot{n}\rangle=-\gamma(\langle n\rangle-\langle
 n\rangle^{eq}) \label{dotn},
 \end{equation}
where $\gamma=W_{-+}+W_{+-}$ and
\begin{equation}
\langle
n\rangle^{eq}=-\frac{1}2\frac{W_{-+}-W_{+-}}{W_{-+}+W_{+-}}=-\frac{1}2\tanh\left(\frac{1}2\beta\hbar\omega\right).
\end{equation}
From Eq. (\ref{dotn}), we find that instantaneous population $\langle n(t)
\rangle$ along the thermalization process (staring at initial time
$t=0$)  can be written in terms of the population $\langle
n(0)\rangle$,
\begin{equation}
 \langle {n}(t)\rangle=\langle n\rangle^{eq}+[\langle
 n(0)\rangle-\langle n\rangle^{eq}]e^{-\gamma t}. \label{ntgt}
\end{equation}
\section{Relation between populations at the begin and the end of a unitary driving process}\label{B}

We consider the unitary time evolution  along the driving
compression $A\rightarrow B$ from $t=0$ to $t=\tau_{ch}$ to
determine $\langle n^2(0)\rangle$ and $\langle
n^2(\tau_{ch})\rangle$ at $A$ and $ B$, respectively. Using
$p_n^0=e^{-\beta_A n\hbar\omega_c}/Z_A$ with
$Z_A=e^{-\beta_A\hbar\omega_c/2}+e^{\beta_A\hbar\omega_c/2}$, it
follows that $\langle n^2(0)\rangle=\sum_{n}n^2p_n^0=1/4$.
 Meanwhile, $\langle n^2(\tau_{ch})\rangle$ can be calculated as
\begin{equation}
\begin{aligned}
\langle n^2(\tau_{ch})\rangle=&\sum_{n,m} n^2 p_{m\rightarrow
n}^{\tau_{ch}}p_m^0(\beta_A)\\
=&\sum_{n,m} n^2|\langle m|U_{\mathrm{com}}|n\rangle|^2p_m^0(\beta_A)\\
=&\frac{1}{4Z_A}\big[e^{-\beta_A\hbar\omega_c/2}\left(|\langle
+|U_{\mathrm{com}}|+\rangle|^2+\langle
+|U_{\mathrm{com}}|-\rangle|^2\right)\\
&+e^{\beta_A\hbar\omega_c/2}\left(|\langle
-|U_{\mathrm{com}}|+\rangle|^2+\langle
-|U_{\mathrm{com}}|-\rangle|^2\right)
\big]\\
=&\langle n^2(0)\rangle\\
=&\frac{1}4,  \label{nb}
\end{aligned}
\end{equation}
and  $\langle n(0)n(\tau_{ch})\rangle$ reads
\begin{equation}
\begin{aligned}
\langle n(0)n(\tau_{ch})\rangle=&\sum_{n,m} nm p_{n\rightarrow
m}^{\tau_{ch}}p_n^0(\beta_A)\\
=&\sum_{n,m} nm|\langle n|U_{\mathrm{com}}|m\rangle|^2p_n^0(\beta_A)\\
=&\frac{1}{4Z_A}\big[e^{-\beta_A\hbar\omega_c/2}\left(|\langle
+|U_{\mathrm{com}}|+\rangle|^2-\langle
+|U_{\mathrm{com}}|-\rangle|^2\right)\\
&+e^{\beta_A\hbar\omega_c/2}\left(-|\langle
-|U_{\mathrm{com}}|+\rangle|^2+\langle
-|U_{\mathrm{com}}|-\rangle|^2\right)
\big]\\
=&\frac{1}4(1-2\xi),  \label{nb}
\end{aligned}
\end{equation}
where $\xi=|\langle \pm|U_{\mathrm{exp}}|\mp\rangle|^2=|\langle
\pm|U_{\mathrm{com}}|\mp\rangle|^2$ and $1-\xi=|\langle
\pm|U_{\mathrm{exp}}|\pm\rangle|^2=|\langle
\pm|U_{\mathrm{com}}|\pm\rangle|^2$ have been used.  For the unitary
expansion $C\rightarrow D$ of the two-level engine, we therefore
have
\begin{equation}
\langle n^2(\tau_{cyc}-\tau_{c})\rangle=\langle
n^2(\tau_{ch}+\tau_h)\rangle=\frac{1}4, \label{nd}
\end{equation}
and
\begin{equation}
\langle
n(\tau_{ch}+\tau_h)n(\tau_{cyc}-\tau_{c})\rangle=\frac{1}4(1-2\xi).
\label{nd}
\end{equation}

Integrating over the probability density function Eq. (\ref{pw1})in Main text,
the first two central moment of quantum work can be calculated as
\begin{equation}
\begin{aligned}
\langle \mathrm{w}\rangle=&\int \mathrm{w} p(\mathrm{w}) d\mathrm{w} \\
=&\int \mathrm{w} d\mathrm{w} \sum_{m,n,i,j}p_{n\rightarrow
m}^{\tau_{ch}} p_n^0p_{i\rightarrow
j}^{\tau_{hc}}p_i^{\tau_{ch}+\tau_h}\delta\{\mathrm{w}-\mathrm{w}[n(0)\rightarrow
m(\tau_{ch});  i(\tau_{ch}+\tau_h)\rightarrow
j(\tau_{cyc}-\tau_c)]\}\\
=&\int \mathrm{w} d\mathrm{w} \sum_{m,n,i,j}p_{n\rightarrow
m}^{\tau_{ch}} p_n^0p_{i\rightarrow
j}^{\tau_{hc}}p_i^{\tau_{ch}+\tau_h}\delta\{\mathrm{w}-[(E_i^h-E_j^c)-(E_m^h-E_n^c)]\}\\
=&\int \mathrm{w} d\mathrm{w} \sum_{m,n,i,j}p_{n\rightarrow
m}^{\tau_{ch}} p_n^0p_{i\rightarrow
j}^{\tau_{hc}}p_i^{\tau_{ch}+\tau_h}\delta\{\mathrm{w}-\hbar[(i\omega_h-j\omega_c)-(m\omega_h-n\omega_c)]\}\\
=&\hbar[(\omega_c-(1-2\xi)\omega_h]\langle
n(0)\rangle+\hbar[(\omega_h-(1-2\xi)\omega_c]\langle
 n(\tau_{ch}+\tau_h)\rangle,  \label{w0}
\end{aligned}
\end{equation}
and
\begin{equation}
\begin{aligned}
\langle \mathrm{w}^2\rangle=&\int \mathrm{w}^2 p(\mathrm{w}) d\mathrm{w} \\
=&\int \mathrm{w}^2 d\mathrm{w} \sum_{m,n,i,j}p_{n\rightarrow
m}^{\tau_{ch}} p_n^0p_{i\rightarrow
j}^{\tau_{hc}}p_i^{\tau_{ch}+\tau_h}\delta\{\mathrm{w}-\mathrm{w}[n(0)\rightarrow
m(\tau_{ch});  i(\tau_{ch}+\tau_h)\rightarrow
j(\tau_{cyc}-\tau_c)]\}\\
=&\int \mathrm{w}^2 d\mathrm{w} \sum_{m,n,i,j}p_{n\rightarrow
m}^{\tau_{ch}} p_n^0p_{i\rightarrow
j}^{\tau_{hc}}p_i^{\tau_{ch}+\tau_h}\delta\{\mathrm{w}-[(E_i^h-E_j^c)-(E_m^h-E_n^c)]\}\\
=&\int \mathrm{w}^2 d\mathrm{w} \sum_{m,n,i,j}p_{n\rightarrow
m}^{\tau_{ch}} p_n^0p_{i\rightarrow
j}^{\tau_{hc}}p_i^{\tau_{ch}+\tau_h}\delta\{\mathrm{w}-\hbar[(i\omega_h-j\omega_c)-(m\omega_h-n\omega_c)]\}\\
=&\hbar^2\omega_h^2 \langle
n^2(\tau_{ch}+\tau_h)\rangle-2\hbar^2\omega_c\omega_h\langle
n(\tau_{ch}+\tau_h)n(\tau_{cyc}-\tau_{c})\rangle
 -2\hbar^2\omega_c\omega_h\langle n(0)n(\tau_{ch})\rangle\\
 &+\hbar^2\omega_c^2 \langle n^2(0)\rangle+2\hbar^2[(\omega_c-(1-2\xi)\omega_h]\langle
n(0)\rangle[(\omega_h-(1-2\xi)\omega_c]\langle n(\tau_{ch}+\tau_h)\rangle\\
&+\hbar^2\omega_c^2 \langle n^2(\tau_{cyc}-\tau_c)\rangle+\hbar^2\omega_h^2 \langle n^2(\tau_{ch})\rangle.  \label{w20}
\end{aligned}
\end{equation}

With the above results, the second moment of stochastic work can be
simplified as
\begin{equation}
\begin{aligned}
\langle
\mathrm{w}^2\rangle=&2\hbar^2\big\{\frac{1}4[\omega_h^2+\omega_c^2-2\omega_c\omega_h
(1-2\xi)]+[\omega_c-(1-2\xi)\omega_h]\\
&\times\langle
n(0)\rangle[\omega_h-(1-2\xi)\omega_c]\langle
 n(\tau_{ch}+\tau_h)\rangle\big\}.
\end{aligned}
\end{equation}

Combining this with Eq. (\ref{w0}), the work fluctuations, $\delta
\mathrm{w}^2=\langle \mathrm{w}^2\rangle-\langle
\mathrm{w}\rangle^2$,  we can obtain Eq. (\ref{deltaw2}) in Main text.

\end{appendices}

 \textbf{Acknowledgements}
We acknowledge the finical support by National Natural Science
Foudantion (Grant Nos. 11875034 and 11505091), and  the Major
Program of Jiangxi Provincial Natural Science Foundation (No.
20161ACB21006). Y.L.M.
acknowledges the financial support from the State Key Programs
of China under Grant (No. 2017YFA0304204).

\end{document}